\documentclass[aps,prd,nofootinbib,superscriptaddress]{revtex4-2}
\usepackage[T1]{fontenc}
\usepackage[utf8]{inputenc} 
\usepackage[english]{babel}
\usepackage[normalem]{ulem}
\usepackage{xcolor}

\usepackage{amsmath,amssymb,amsfonts}
\usepackage{bm}
\usepackage{latexsym}

\usepackage{graphicx}
\usepackage{array}
\usepackage{booktabs}

\newcommand{\be}{\begin{equation}}
\newcommand{\ee}{\end{equation}}

\usepackage[
  pdfusetitle,
  bookmarks=true,
  bookmarksnumbered=false,
  bookmarksopen=false,
  breaklinks=false,
  pdfborder={0 0 1},
  backref=false,
  colorlinks=false
]{hyperref}

\begin{document}

\title{Perturbative Analytical Construction of Bowen--York Initial Data
for Binary Black Holes up to Third Order in Spin and Linear Momentum}

\author{Leyla Ogurol}
\email{leyla.ogurol@metu.edu.tr}
\affiliation{Department of Physics, Middle East Technical University, 06800, Ankara, Turkey}
\affiliation{Department of Physics, Gazi University, 06500, Ankara, Turkey}

\author{Tore Boybeyi}
\email{boybe001@umn.edu}
\affiliation{School of Physics and Astronomy, University of Minnesota, 55455 MN, USA}

\author{Bayram Tekin}
\email{bayram.tekin@bilkent.edu.tr}
\affiliation{Department of Physics, Bilkent University, 06800 Ankara, Turkey}

\date{\today}

\begin{abstract}
We analytically construct initial data for binary black hole systems by solving the Einstein constraint equations within the Bowen–York framework. Allowing for arbitrarily oriented linear momenta and spins, we obtain a perturbative solution of the vacuum Hamiltonian constraint up to third order in the source parameters. Our solution explicitly incorporates the system's binary nature via position-weighted multipole moments. The momentum constraints are solved exactly, while the conformal factor is determined analytically. Using the resulting initial data, we compute the ADM energy, linear and angular momenta, and the irreducible mass. We also determine the shape of the apparent horizon and show that it is geometrically consistent with the inner-end regularity. The resulting expressions are written in a manifestly tensorial form and provide an analytic benchmark for numerical-relativity calculations within the perturbative far-zone regime considered here.
\end{abstract}

\maketitle
\tableofcontents

\section{Introduction}

The field of astrophysics has been profoundly transformed by the direct detection of
gravitational waves, first achieved by the Laser Interferometer Gravitational Wave Observatory (LIGO) \cite{abbott2016ligo,abbott2016observation}.
This breakthrough confirmed a central prediction of Einstein’s theory of General Relativity, opening a new observational window onto the Universe.
More recently, the North American Nanohertz Observatory for Gravitational Waves
(NANOGrav) has reported evidence for a stochastic Gravitational Wave background
\cite{McLaughlin:2013ira}, further expanding the range of accessible Gravitational
Wave sources. Together, these developments have ushered in an era in which strong-field gravity can be
probed directly through observations, with binary black hole systems playing a central
role as some of the most powerful emitters of gravitational radiation.

Binary black hole systems undergo a rich dynamical evolution: two compact objects inspiral
under gravitational radiation reaction, merge into a highly distorted remnant, and
eventually settle into a stationary Kerr black hole through the emission of quasinormal
modes \cite{Ferrari_2008}.
Accurate modeling of the gravitational waves emitted during these phases requires the use of numerical relativity, in which Einstein’s equations are solved using fully nonlinear
computational methods.
A major milestone in this field was achieved by Pretorius in 2005, who performed the first
successful simulation of a binary black hole merger, overcoming longstanding numerical
instabilities \cite{pretorius2005gq}.
Since then, numerical relativity has become a mature and indispensable tool in
gravitational-wave physics.

The generation of gravitational waves during the inspiral, merger, and ringdown phases
has been extensively studied using numerical relativity
\cite{Centrella:2010mx,Baumgarte:2002jm}, particularly for the merger regime, where the strongest fields and highest frequencies are produced.
While the evolution equations of General Relativity have been explored in great detail, comparatively less attention has been paid to the structure and solution of the constraint equations \cite{Bartnik:2002cw}, which determine the allowed initial data on a spatial
hypersurface. For binary black hole systems—especially in configurations involving momenta, spins,
or close separations—the construction of physically meaningful and analytically controlled initial data remains a challenging problem.

In this work, we build on and extend previous analytical studies of the constraint equations
\cite{altas2023perturbative,Dennison:2006nq} by developing a perturbative solution for
binary black holes with \emph{arbitrary} linear momenta and spins.
In particular, we relax the simplifying assumptions on the relative orientations of the spins
and momenta adopted in earlier treatments.
Achieving this requires substantial analytical work, including the systematic decoupling
of partial differential equations involving tensorial angular structures.
Within this framework, we construct explicit solutions up to third order in the source
parameters and perform a number of internal consistency checks.
The methods developed here are fairly general and can, in principle, be extended to higher
orders and to systems involving more than two black holes.
Such multi–black hole interactions may become increasingly relevant, as suggested by recent
studies \cite{Bamber_2025}, and could provide insight into the formation of black holes with
masses of order $100\,M_\odot$ \cite{LIGOScientific:2025rsn}.

The central advance of this work is twofold. Methodologically,
we show that lifting the orientation restrictions of earlier analytic treatments
turns the constraint problem from an effectively axisymmetric one into a fully
tensorial elliptic system, and we solve it in closed form by a systematic
block diagonalization in tensor spherical harmonics carried to third order in the
source parameters. Physically, this generality is what first exposes the
spin--orbit sector of the initial data: the leading orientation-dependent
contribution to the ADM energy, absent by symmetry in all previous analytic
constructions, appears here as a genuine third-order effect. Together these make
the present data the first analytic Bowen--York initial data applicable to
generically misaligned, precessing binaries.

The paper is organized as follows.
In Sec.~II, we review the initial value problem of General Relativity and the associated
constraint equations. In Sec.~III, we introduce the notation and conventions used throughout this work.
In Sec.~IV, we solve the momentum constraints and derive the nonlinear elliptic equation
arising from the Hamiltonian constraint, along with expressions for the total linear and
angular momentum and the ADM mass.
In Sec.~V, we introduce the perturbative expansion used to solve the constraint equations.
Higher-order solutions are developed in the same section and applied explicitly in Sec.~VI, where we
also compute the ADM and irreducible masses.
Finally, in Sec.~VII, we determine the location and shape of the apparent horizon.

Throughout, the construction rests on a few assumptions that we state here at the outset and develop where they are used: the initial data are conformally flat, maximally sliced, and in vacuum, as is standard in the Bowen–York approach, and the linear momenta and spins are treated perturbatively, so that all derived expressions hold to a fixed order in these parameters. The closed-form conformal factor and the ADM energy are obtained in the far zone, at separations large compared to the binary separation, whereas the inner-end masses and the apparent-horizon geometry are extracted from the near-puncture asymptotics rather than by extrapolating the far-zone expansion inward. The explicit expressions for the ADM and irreducible masses are quoted for equal-mass binaries, and the horizon analysis assumes that the two black holes are separated widely enough that no common apparent horizon has formed.

\section{The Initial Value Problem of General Relativity}
\label{sec:IVP}

Numerical relativity aims to model a spacetime geometry described by a Lorentzian metric
$g_{\mu\nu}$ $(\mu,\nu=0,1,2,3)$ that satisfies Einstein’s field equations,
\begin{equation}
G_{\mu\nu}=8\pi T_{\mu\nu},
\end{equation}
where we work in units $G=c=1$.
The principal difficulties arise from the nonlinearity of the equations and their
diffeomorphism (gauge) invariance.

In the standard $3{+}1$ decomposition
\cite{Arnowitt:1962hi,york1979}, the spacetime is assumed to be globally hyperbolic and foliated by spacelike hypersurfaces
$\Sigma_t$ labeled by a time function $t$.
Each hypersurface admits a future-directed timelike unit normal $n^\mu$.
The induced spatial metric on $\Sigma_t$ is
\begin{equation}
\gamma_{\mu\nu}:=g_{\mu\nu}+n_\mu n_\nu,
\end{equation}
which constitutes half of the canonical coordinates in phase space.
The conjugate momenta are related to the extrinsic curvature:
\begin{equation}
K_{\mu\nu}:=-\tfrac{1}{2}\mathcal{L}_n\gamma_{\mu\nu},
\end{equation}
where $\mathcal{L}_n$ denotes the Lie derivative along $n^\mu$.
Introducing coordinates $(t,x^i)$ adapted to the foliation, with $i,j,\ldots=1,2,3$,
the purely spatial components of these tensors are denoted by $\gamma_{ij}$ and $K_{ij}$.

With these definitions, the spacetime line element takes the form
\begin{equation}
ds^2=-\alpha^2 dt^2+\gamma_{ij}\bigl(dx^i+\beta^i dt\bigr)
                         \bigl(dx^j+\beta^j dt\bigr),
\end{equation}
where $\alpha$ is the lapse function and $\beta^i$ the shift vector, both functions of time and space.
The lapse function determines the proper time separation between adjacent hypersurfaces along
$n^\mu$, while the shift vector encodes the displacement of spatial coordinates from one slice to
the next.
The time-evolution vector field is
$t^\mu=\alpha n^\mu+\beta^\mu$, with $\beta^\mu n_\mu=0$.

When Einstein’s equations are evaluated on a hypersurface $\Sigma_t$, they split into
constraint and evolution equations.
The constraint equations consist of one Hamiltonian constraint and three momentum constraints,
\begin{align}
\Phi_0 &:= {}^{\Sigma}\!R + K^2 - K_{ij}K^{ij} -16\pi\rho = 0,
\label{Hamiltonianconstraint}\\
\Phi^i &:= D_j\!\left(K^{ij}-\gamma^{ij}K\right)-8\pi Y^i = 0,
\label{momentumconstraint}
\end{align}
see, e.g., \cite{baumgarte2010numerical}.
Here ${}^{\Sigma}\!R$ is the scalar curvature of the hypersurface $(\Sigma_t,\gamma_{ij})$,
$K:=\gamma^{ij}K_{ij}$ is the mean extrinsic curvature,
and $D_i$ is the Levi-Civita connection compatible with $\gamma_{ij}$.
The matter source terms are defined by
\begin{align}
\rho &:= n^\mu n^\nu T_{\mu\nu}, \nonumber\\
Y^i &:= -\gamma^{i\nu} n^\mu T_{\nu\mu}.
\end{align}

Although the present work focuses on the constraint equations, it is useful to recall that
they are supplemented by evolution equations for $(\gamma_{ij},K_{ij})$.
Rather than writing these explicitly, we record their compact Fischer--Marsden form
\cite{Altas:2019kci},
\begin{equation}
\frac{d}{dt}
\begin{pmatrix}
\gamma \\
\pi
\end{pmatrix}
=
J \circ D\Phi^*(\gamma,\pi)(N),
\qquad
J :=
\begin{pmatrix}
0 & 1 \\
-1 & 0
\end{pmatrix},
\label{eq:15}
\end{equation}
where $\pi$ denotes the canonical momentum conjugate to $\gamma$,
$N:=(\alpha,\beta^i)$ is the lapse--shift pair,
$\Phi(\gamma,\pi):=(\Phi_0,\Phi^i)$ is the constraint map,
and $D\Phi^*$ is the adjoint of its linearization.
Together, Eqs.~\eqref{Hamiltonianconstraint}, \eqref{momentumconstraint}, and
\eqref{eq:15} are equivalent to Einstein’s equations under the assumption of global
hyperbolicity.
They also make explicit the constrained Hamiltonian structure of General Relativity; see
\cite{Gourgoulhon:2012ffd} for a detailed discussion.

Numerical relativity simulations require the construction of initial data
$(\gamma_{ij},K_{ij})$ that satisfy the constraint equations.
While the constraints restrict the allowed data, they do not uniquely determine which
components are fixed, and which remain freely specifiable.
In the linearized theory, one can cleanly separate dynamical, constrained, and gauge
degrees of freedom, but no unique decomposition exists in the full nonlinear theory.
Consequently, one must adopt a specific scheme that recasts the constraint equations into
an elliptic system with suitable boundary conditions
\cite{OMurchadha:1973byk,OMurchadha:1974xtz,baumgarte2010numerical}.

A widely used approach is the conformal transverse–traceless (CTT), or
York–Lichnerowicz, decomposition
\cite{bowen1980,PhysRevLett.26.1656,PhysRevLett.28.1082,JMPA_1944_9_23__37_0}.
Although CTT is not unique, it provides a natural framework for constructing analytic and
numerical initial data.
An alternative is the conformal thin-sandwich (CTS) decomposition
\cite{york1999CTS,pfeiffer2003extrinsic}, which is particularly well suited to
quasi-equilibrium configurations.
In this work, we adopt the CTT framework and specialize it in the following section to
Bowen–York initial data. However, we first start with the notations and conventions we shall use. 

\section{Notation and Conventions}
\label{sec:notation}

To avoid ambiguity and make the text easier to read, we lay out our notation and conventions here. 

Greek indices $\mu,\nu,\ldots$ denote spacetime tensors, while Latin indices
$i,j,k,\ldots$ denote tensors intrinsic to the initial-data hypersurface $\Sigma$.

\begin{itemize}
\item $g_{\mu\nu}$ is the spacetime metric, with the  Levi-Civita connection $\nabla_\mu$.

\item $\Sigma$ is a spacelike hypersurface with the induced 3-metric $\gamma_{ij}$ and
the compatible covariant derivative $D_i$.

\item $n^\mu$ is the \emph{future-directed timelike unit normal} to $\Sigma$:
\begin{equation}
n^\mu n_\mu=-1,\qquad n^\mu \gamma_{\mu\nu}=0.
\end{equation}

\item $K_{ij}$ is the physical extrinsic curvature of $\Sigma$,
\begin{equation}
K_{ij} = -\gamma_i{}^{\mu}\gamma_j{}^{\nu}\nabla_\mu n_\nu,
\end{equation}
and throughout this work, we adopt the standard Bowen--York/maximal condition
$K:=\gamma^{ij}K_{ij}=0$.
\end{itemize}

We employ conformal flatness,
\begin{equation}
\gamma_{ij}=\psi^{4}\,\eta_{ij},
\end{equation}
where $\eta_{ij}$ is the flat metric on $\mathbb{R}^3$, written in Cartesian or spherical
coordinates as appropriate.
Quantities with hats refer to objects defined with respect to the conformal (flat)
geometry:
\begin{equation}
K_{ij}=\psi^{-2}\,\hat K_{ij},\qquad
D_i \text{ is compatible with } \gamma_{ij},\qquad
\hat D_i \text{ is compatible with } \eta_{ij}.
\end{equation}

Angular dependence is parameterized by $\Omega=(\theta,\phi)$ and is equivalently described by the
unit radial vector
\begin{equation}\label{unitradial}
\hat n(\theta,\phi)=
(\sin\theta\cos\phi,\ \sin\theta\sin\phi,\ \cos\theta).
\end{equation}
The Laplacian on the unit sphere is
\begin{equation} \label{Laplacian}
\Delta := \partial^2_\theta + \cot\theta\,\partial_\theta
+ \frac{1}{\sin^2\theta}\,\partial^2_\phi.
\end{equation}

In Sec.~\ref{sec:AH}, the symbol $s_i$ denotes the \emph{outward unit normal} to the
apparent-horizon surface $\mathcal{S}\subset\Sigma$ \emph{within the spatial geometry}
$(\Sigma,\gamma_{ij})$.
The induced metric on $\mathcal{S}$ is $q_{ij}=\gamma_{ij}-s_i s_j$.
We also use the conformally rescaled normal $\hat s^i=\psi^{-2} s^i$.

A bookkeeping parameter $\epsilon$ (set to unity at the end) is used to label
\emph{source order} in the Bowen--York data, i.e.\ powers of linear momentum, spin, and
their position-weighted moments. Independently, the \emph{far-zone} behavior is organized in powers of $1/r$
(or $\kappa=a/r$).
A given term may therefore be, for example, second order in source parameters while
leading order in $1/r$; both orderings are tracked explicitly in Sec.~\ref{sec:IV}.

\section{The Einstein Constraint Equations}
\label{sec:constraints}

The conformal transverse--traceless (CTT) approach begins with a conformal transformation of
the spatial metric,
\begin{equation}
\gamma_{ij}=\psi^{4}\hat\gamma_{ij},
\end{equation}
where $\psi>0$ is the conformal factor and $\hat\gamma_{ij}$ is the conformally related
metric.
(Throughout, hatted quantities are defined with respect to $\hat\gamma_{ij}$.)

It is useful to decompose the extrinsic curvature into its trace $K$ and traceless part
$A_{ij}$,
\begin{equation}
K_{ij}=A_{ij}+\frac{1}{3}K\,\gamma_{ij}.
\label{extrinsica}
\end{equation}
In the CTT scheme, one adopts the conformal scalings
\begin{equation}
A^{ij}=\psi^{-10}\hat A^{ij},\qquad K=\hat K,
\end{equation}
so that, in particular, $A^{ij}$ is divergence-free if and only if $\hat A^{ij}$ is.
The scalar curvature transforms as
\begin{equation}
R=\psi^{-4}\hat R-8\psi^{-5}\hat D^2\psi,
\end{equation}
where $\hat D^2:=\hat\gamma^{ij}\hat D_i\hat D_j$ is the Laplacian associated with
$\hat\gamma_{ij}$.
With these choices, the Hamiltonian constraint \eqref{Hamiltonianconstraint} and momentum
constraints \eqref{momentumconstraint} become
\begin{equation}
\begin{aligned}
8\hat D^2\psi-\psi\hat R-\frac{2}{3}\psi^5 K^2+\psi^{-7}\hat A_{ij}\hat A^{ij}
&=-16\pi \psi^5\rho, \\
\hat D_j\hat A^{ij}-\frac{2}{3}\psi^6\hat\gamma^{ij}\hat D_j K
&=8\pi \psi^{10}Y^i .
\end{aligned}
\label{einsteinconstrainttwo}
\end{equation}

In this work, we specialize to conformal flatness and maximal slicing,
\begin{equation}
\hat\gamma_{ij}=\eta_{ij},\qquad K=0,
\end{equation}
and to vacuum initial data $\rho=Y^i=0$.
Then \eqref{einsteinconstrainttwo} reduces to
\begin{align}
\hat D_i\hat D^i\psi &= -\frac{1}{8}\psi^{-7}\hat A_{ij}\hat A^{ij},
\label{hamiltonian}\\
\hat D_j\hat A^{ij} &= 0,
\label{momentum}
\end{align}
where $\hat D_i$ is now the flat covariant derivative.

Any symmetric trace-free tensor admits the (York) decomposition into a transverse-traceless
part and a longitudinal part generated by a vector potential:
\begin{equation}
\hat A^{ij}=\hat A^{ij}_{TT}+\hat A^{ij}_L,
\qquad
\hat D_j\hat A^{ij}_{TT}=0,
\end{equation}
with
\begin{equation}
\hat A^{ij}_L
=\hat D^i W^j+\hat D^j W^i-\frac{2}{3}\hat\gamma^{ij}\hat D_k W^k
\equiv (\hat L W)^{ij},
\end{equation}
where $(\hat L W)^{ij}$ is symmetric and trace-free by construction.
In what follows, we set $\hat A^{ij}_{TT}=0$ and solve the momentum constraint using the
Bowen–York prescription \cite{bowen1980}.
Accordingly, we will denote the resulting conformal extrinsic curvature by $\hat K_{ij}$
(so that $\hat A_{ij}\equiv \hat K_{ij}$ in the present maximal setting).

\subsection{Bowen--York solution of the momentum constraint for two punctures}
\label{sec:BY}

Because \eqref{momentum} is linear, it admits exact analytic solutions and allows for superposition.
Consider two punctures located at coordinate positions $\vec c_A$ ($A=1,2$) that carry individual linear momentum $p_A^{\,i}$ and spin (intrinsic angular
momentum) $j_A^{\,i}$. The Bowen-York conformal extrinsic curvature for puncture $A$ is the sum of a momentum part
and a spin part, and the binary solution is their superposition:
\begin{equation}
\hat K_{ij}
=
\hat K^{(P)}_{ij}(\vec c_1,\vec p_1)
+\hat K^{(J)}_{ij}(\vec c_1,\vec j_1)
+\hat K^{(P)}_{ij}(\vec c_2,\vec p_2)
+\hat K^{(J)}_{ij}(\vec c_2,\vec j_2).
\label{eq:BY_superposition}
\end{equation}
 Defining the coordinates of the punctures and the unit normal as 
\begin{equation}\label{punc}
r_A := |\vec x-\vec c_A|,\qquad
\hat n_A^{\,i}:=\frac{x^i-c_A^{\,i}}{r_A},
\end{equation}
one has 
\begin{align}
\hat K_{ij}
&=
\frac{3}{2r_1^{2}}
\Big(2p_{1(i}\hat n_{1j)}+(\hat n_{1i} \hat n_{1j}-\eta_{ij})\,p_1\!\cdot \hat n_1\Big)
+\frac{3}{r_1^{3}}(j_1\times \hat n_1)_{(i}\hat n_{1j)} +\ (1\rightarrow 2),
\label{extrinsic}
\end{align}
where symmetrization is $a_{(i}b_{j)}=(a_ib_j+a_jb_i)/2$, and all contractions/cross products
are taken with respect to the flat metric $\eta_{ij}$.
Equation \eqref{extrinsic} solves \eqref{momentum} exactly for all $\vec x\neq \vec c_A$.
The construction, therefore, reduces the problem to solving the nonlinear Hamiltonian
constraint \eqref{hamiltonian} for the conformal factor $\psi$.

\subsection{Conserved quantities at spatial infinity}

Even before solving \eqref{hamiltonian}, the total ADM linear and angular momenta can be
computed from the asymptotic behavior of the extrinsic curvature.
For asymptotically flat data,
\begin{equation}
\psi(r)=1+\frac{E_{\rm ADM}}{2r}+\mathcal{O}\!\left(\frac{1}{r^{2}}\right),
\qquad r\to\infty,
\label{eq:psi_asymptotics}
\end{equation}
and the ADM energy $E_{\rm ADM}$ is determined by the $1/r$ coefficient (equivalently, by the
surface integral of $\partial_i\psi$ at infinity), which will be evaluated once the
perturbative solution for $\psi$ is obtained.

Define $dS_i=\hat n_i\,dS$ as the surface element on $S^2_\infty$, with $\hat n_i$ as the outward unit
normal to the large coordinate spheres.
The ADM linear momentum can be written as \cite{Baumgarte:2002jm}
\begin{equation}
P^{i}=\frac{1}{8\pi}\int_{S_{\infty}^{2}} K^{ij}\, dS_j
     =\frac{1}{8\pi}\int_{S_{\infty}^{2}} \hat K^{ij}\, dS_j,
\end{equation}
and the total ADM angular momentum as
\begin{equation}
J_{i}
=\frac{1}{8\pi}\,\epsilon_{ijk}
\int_{S_{\infty}^{2}} x^{j}\,K^{kl}\,dS_l
=\frac{1}{8\pi}\,\varepsilon_{ijk}
\int_{S_{\infty}^{2}} x^{j}\,\hat K^{kl}\,dS_l.
\label{ang..}
\end{equation}
(Equivalently, in the absence of axisymmetry, one may use the rotational Killing vectors
$\xi^{n}_{(i)}=\epsilon^{n}{}_{mi}x^m$ in the standard Komar/ADM expression.)

For the binary Bowen–York solution \eqref{extrinsic}, these integrals yield
\begin{align}
P_i &= p_{1i}+p_{2i},
\label{linearmomentum}\\
J_i &= j_{1i}+j_{2i}.
\label{angularmomentum}
\end{align}

Finally, the ADM energy can be expressed in terms of the deviation
$h_{ij}:=(\psi^{4}-1)\delta_{ij}$ from flat space as
\begin{align}
E_{\mathrm{ADM}}
&=\frac{1}{16\pi}\int_{S_\infty^2} dS_i
\left(\partial_j h^{ij}-\partial^i h^j{}_{j}\right)
= -\frac{1}{2\pi}\int_{S_\infty^2} dS_i\,\partial_i\psi,
\label{ADM}
\end{align}
which we evaluate after determining $\psi$ perturbatively.

\section{Perturbative Scheme and Far-Zone Expansion}
\label{sec:IV}

We now solve the nonlinear Hamiltonian constraint
\begin{equation}
\hat D^i \hat D_i \psi = -\frac{1}{8}\,\psi^{-7}\,\hat K_{ij}\hat K^{ij},
\label{eq:HC}
\end{equation}
for Bowen–York binary black hole initial data.
The momentum constraint has already been solved exactly by the superposition of a single-puncture
solution, yielding a genuine two-puncture extrinsic curvature $\hat K_{ij}(\vec x)$.
The remaining difficulty is the nonlinear elliptic equation \eqref{eq:HC}, which we solve
\emph{perturbatively} and \emph{analytically} by combining (i) a source-parameter expansion
and (ii) a systematic far-zone ($1/r$) expansion.

\subsection{Scales, coordinates, and regimes}
\label{sec:IV.A}

We consider two punctures at fixed coordinate positions $\vec c_1$ and $\vec c_2$ in the
conformally flat background.
For a field point $\vec x\in\mathbb{R}^3$, define the puncture-centered radii as before (\ref{punc}).

For asymptotic and multipolar expansions, we introduce a \emph{global} radial
coordinate with respect to a chosen origin (conveniently taken as the center of mass),
\begin{equation}
r := |\vec x|,\qquad \hat n := \frac{\vec x}{r}.
\label{eq:r_n_def}
\end{equation}
Here $\hat n$ is the outward unit normal on coordinate spheres $S^2_r$ at large $r$.
This $\hat n$ is \emph{not} the 3+1 timelike normal $n^\mu$ used in Sec.~\ref{sec:IVP}, nor is it
$\hat n^i_A$; it is purely the spatial unit radial direction at infinity.

Our closed-form analytic solution for the Hamiltonian constraint is constructed in the
\emph{far--zone} region
\begin{equation}
r \gg |\vec c_1|,\ |\vec c_2|\qquad\text{(equivalently: } r \gg d,\ \ d:=|\vec c_1-\vec c_2| \text{)}.
\label{eq:farzone_regime}
\end{equation}
In this regime, the puncture radii admit expansion
\begin{equation}
\frac{1}{r_A}
= \frac{1}{r}
+ \frac{\hat n\cdot \vec c_A}{r^2}
+ \frac{1}{2r^3}\Big(3(\hat n\cdot\vec c_A)^2-c_A^2\Big)
+ \mathcal{O}(r^{-4}),
\label{eq:1overrA_expansion}
\end{equation}
and similarly for products involving $ n^i_A$.
It is important to clarify the choice of coordinates. While the Bowen-York extrinsic curvature is exact for two punctures at arbitrary locations $\vec{c}_A$, solving the Hamiltonian constraint analytically requires a global coordinate expansion. In this section, $r$ denotes the radial coordinate from the center of mass. The binary geometry is encoded in the multipole moments ($P, J, M, L$) defined below, which depend explicitly on the puncture positions $\vec{c}_1$ and $\vec{c}_2$.

Therefore, any two-puncture quantity (such as $\hat K_{ij}$ and $\hat K_{ij}\hat K^{ij}$)
admits a systematic multipole expansion in inverse powers of $r$, with coefficients built
from the binary parameters.

A key point is that we keep track of two distinct kinds of ``order'':
\begin{itemize}
\item \textbf{Source order:} the perturbative order in the Bowen-York source parameters
(linear momenta, spins, and position-weighted moments).
\item \textbf{Far-zone order:} the falloff order in inverse powers of $r$.
\end{itemize}
A term may be, for example, second order in the source parameters but leading order in
$1/r$.

In addition to the total ADM momentum and spin (computed from the exact momentum-constraint
solution),
we define the following position-weighted moments:
\begin{align}
M_{ij} &:= p_{1i}c_{1j}+p_{2i}c_{2j}, \nonumber\\
L_{ij} &:= M_{ij}+M_{ji}, \nonumber\\
N_{ij} &:= j_{1i}c_{1j}+j_{2i}c_{2j}, \nonumber\\
O_{ij} &:= N_{ij}+N_{ji}.
\label{eq:MNLO_defs}
\end{align}
These objects naturally appear as coefficients of the $1/r^3$ and higher multipole
structures of the far--zone fields.

\subsection{Perturbative ansatz and hierarchy}
\label{sec:IV.ansatz}

Starting from the exact two-puncture Bowen-York solution and using the far--zone expansion
\eqref{eq:1overrA_expansion}, the extrinsic curvature admits the large-$r$ representation
\begin{align}
\hat K_{ij}
&= \frac{3}{2r^2}\Big( (P\cdot \hat n)\,(\hat n_i\hat n_j-\eta_{ij}) + 2P_{(i}\hat n_{j)}\Big)
\nonumber\\
&
-\frac{3}{2r^3}\Big(
L_{ij}
-6M_{j(k}\hat n^k \hat n_{i)}
+2M_{k(j}\hat n^k \hat n_{i)}
-4(J\times \hat n)_{(i}\hat n_{j)}
+M^k{}_k(\hat n_i\hat n_j-\eta_{ij})
\nonumber\\
&\qquad \quad
+M_{kl}\hat n^k\hat n^l\,(3\eta_{ij}-5\hat n_i\hat n_j)
\Big)
+\mathcal{O}(r^{-4}).
\label{eq:Kij_farzone}
\end{align}
Here $P_i$ and $J_i$ are the total ADM linear and angular momenta, and $\eta_{ij}$ is the
flat metric.
Equation \eqref{eq:Kij_farzone} is \emph{not} a single-puncture replacement of the binary
data; it is the \emph{far--zone expansion} of the genuine two-puncture configuration.

We expand the conformal factor as a perturbation series in the source parameters:
\begin{equation}
\psi(r,\Omega)
=\psi_0(r)+\psi_1(r,\Omega)+\psi_2(r,\Omega)+\psi_3(r,\Omega)+\cdots,
\label{eq:psi_series}
\end{equation}
where $\psi_0$ denotes the contribution at zeroth order, $\psi_1$ first order, and so on in the source parameters.
The angular dependence is encoded either by $\Omega=(\theta,\phi)$, or equivalently by the
unit vector $\hat n(\theta,\phi)$.
Asymptotic flatness at the outer end requires
\begin{equation}
\psi(r,\Omega)\to 1 \qquad (r\to\infty),
\label{eq:BC_infty}
\end{equation}
and puncture data admit the standard near-puncture singular behavior
\begin{equation}
\psi(r,\Omega) \sim 1+\frac{a}{r} \qquad (r\to 0),
\label{eq:BC_puncture}
\end{equation}
with bounded remainder (made precise in Sec.~\ref{sec:V}).
Correspondingly, we impose
\begin{equation}
\psi_0(r)=1+\frac{a}{r},\qquad
\psi_n(r,\Omega)\to 0 \ \ (r\to\infty)\quad \text{for } n\ge 1.
\label{eq:BC_split}
\end{equation}

At zeroth order, \eqref{eq:HC} reduces to Laplace's equation:
\begin{equation}
\hat D^i\hat D_i \psi_0 = 0,
\end{equation}
with solution $\psi_0=1+a/r$ consistent with \eqref{eq:BC_split}.
At first order, there is no scalar source because $\hat K_{ij}\hat K^{ij}$ is quadratic in
the Bowen–York data; hence,
\begin{equation}
\hat D^i\hat D_i \psi_1 = 0,\qquad \psi_1=0,
\label{eq:psi1_zero}
\end{equation}
given the decay condition at infinity.

At the second and third orders, the Hamiltonian constraint yields
\begin{align}
\hat D^i\hat D_i \psi_2
&= -\frac{1}{8}\,\psi_0^{-7}\,\big(\hat K_{ij}\hat K^{ij}\big)^{(2)},
\label{eq:psi2_eq}\\
\hat D^i\hat D_i \psi_3
&= -\frac{1}{8}\,\psi_0^{-7}\,\big(\hat K_{ij}\hat K^{ij}\big)^{(3)},
\label{eq:psi3_eq}
\end{align}
where the superscript $(n)$ denotes the part of $\hat K_{ij}\hat K^{ij}$ that is $n$th
order in the source parameters.
At these orders, it is sufficient to substitute $\psi\to\psi_0$ on the right-hand side 
because any correction $\psi=\psi_0+\psi_2+\cdots$ would contribute only at higher source
order.

\subsection{Block-diagonalization of the angular operators}
\label{sectionA}

We now explain how to solve equations of the form \eqref{eq:psi2_eq}--\eqref{eq:psi3_eq} once the source is expressed as a sum of angular tensor structures.
Consider the general Poisson equation
\begin{align}
\hat D^{i}\hat D_{i}\,\zeta
= \Big(\frac{\partial^{2}}{\partial r^{2}}
+\frac{2}{r}\frac{\partial}{\partial r}
+\frac{\Delta}{r^{2}}\Big)\zeta
= \xi^{i_{1}\!\cdots i_{s}}(\hat n)\,S_{i_{1}\!\cdots i_{s}}(r),
\label{eq:general_poisson}
\end{align}
where the Laplacian on the sphere is given in (\ref{Laplacian})
and $(S_{i_{1}\cdots i_{s}}(r))$ is independent of \(\hat n\), denoting the unit radial direction vector expressed in terms of the angles as (\ref{unitradial}).
We expand $\xi^{i_{1}\!\cdots i_{s}}$ in eigen-tensor representatives
$A^{i_{1}\!\cdots i_{s}}_{(\ell)}(\hat n)$ (see Table~\ref{tab:angular-basis}), which satisfy
\begin{align}
  &\Delta A^{i_{1}\!\cdots i_{s}}_{(\ell)}(\hat n)
    =-\,\ell(\ell{+}1)\,A^{i_{1}\!\cdots i_{s}}_{(\ell)}(\hat n), \nonumber \\
  &A^{i_{1}\!\cdots i_{s}}_{(k)}(\hat n)\,
  A_{(\ell)\,i_{1}\!\cdots i_{s}}(\hat n)
    =\delta_{k\ell}.
\end{align}
Hence, one has the expansion 
\begin{align}
  \xi^{i_{1}\!\cdots i_{s}}(\hat n)
  = \sum_{\ell} c_{(\ell)}\,A^{i_{1}\!\cdots i_{s}}_{(\ell)}(\hat n),
\end{align}
where $c_{(\ell)}$ is given as
\begin{align}
  c_{(\ell)}=\xi^{i_{1}\!\cdots i_{s}} A_{(\ell)\,i_{1}\!\cdots i_{s}}
\end{align}
by orthonormality.  Then we have the decomposition
\begin{align}
  \zeta(r,\hat n)
  = \sum_{\ell} A^{i_{1}\!\cdots i_{s}}_{(\ell)}(\hat n)\,
      \zeta^{(\ell)}_{i_{1}\!\cdots i_{s}}(r),
\end{align}
which yields independent, decoupled, radial equations
\begin{align}\label{eq:radial_ode}
  \Big(\frac{d^{2}}{dr^{2}}+\frac{2}{r}\frac{d}{dr}
      -\frac{\ell(\ell+1)}{r^{2}}\Big)
  \zeta^{(\ell)}_{i_{1}\!\cdots i_{s}}(r)
  = c_{(\ell)}\,S_{i_{1}\!\cdots i_{s}}(r),
\end{align}
with the boundary conditions consistent with \eqref{eq:BC_split}. This projection diagonalizes
the angular operator (block-diagonal in $\ell$).

To make the projection-and-solve procedure fully explicit, we carry the
$PP$ sector as an example. The source of (\ref{eq:PP}) is a sum of an
$\ell=0$ and an $\ell=2$ eigentensor of Table~\ref{tab:angular-basis}, with
coefficients read off from the decomposition (\ref{eq:PPdecomp}). The two
sectors decouple, giving two radial equations of the form (\ref{eq:radial_ode}),
\begin{align}
\left(\frac{d^{2}}{dr^{2}}+\frac{2}{r}\frac{d}{dr}\right)
   \zeta^{(0)}(r)
   &= c_{(0)}\,S(r), \\[2pt]
\left(\frac{d^{2}}{dr^{2}}+\frac{2}{r}\frac{d}{dr}
      -\frac{6}{r^{2}}\right)\zeta^{(2)}(r)
   &= c_{(2)}\,S(r),
\end{align}
with the common radial source
$S(r)=-\,9\kappa^{4}/\big[16a^{2}(1+\kappa)^{7}\big]$ and the angular
coefficients $c_{(0)}=5/\sqrt3$, $c_{(2)}=-2\sqrt2/\sqrt3$ following from
(\ref{eq:PPdecomp}). Each is a linear inhomogeneous Euler-type equation;
imposing regularity together with the decay conditions (\ref{eq:BC_split})
($\zeta^{(\ell)}\to 0$ as $r\to\infty$) fixes the homogeneous
coefficients uniquely. Reassembling
$\psi_2^{PP}=A_{(0)}\zeta^{(0)}+A_{(2)ij}\hat n^i\hat n^j\zeta^{(2)}$
with the Table~\ref{tab:angular-basis} representatives reproduces
(\ref{eq:psiPPapp}), i.e. the closed form quoted in the Appendix. The
remaining sectors ($JJ$, $PJ$, and the third-order structures) are
treated identically, the only change being which $\ell$ blocks are
populated, as dictated by Table~\ref{tab:angular-basis}.

\small
\begin{table*}[t]
\centering
\caption{Eigentensor representatives $A^{\cdots}_{(\ell)}(\hat n)$ used to diagonalize $\Delta$. Normalization:
$A^{i_{1}\cdots i_{s}}_{(k)}A_{(\ell)\,i_{1}\cdots i_{s}}=\delta_{k\ell}$.}
\vspace{0.25em}
\resizebox{\textwidth}{!}{%
\renewcommand\arraystretch{1.2}
\begin{tabular}{|l|c|c|c|c|}
\hline
$\boldsymbol{V}^{\otimes n}$ & $\ell=0$ & $\ell=1$ & $\ell=2$ & $\ell=3$ \\ \hline
$n=1$ & -- & $\hat n^{i}$ & -- & -- \\ \hline
$n=2$ &
  $\tfrac{1}{\sqrt{3}}\,\eta^{ij}$ &
  $\tfrac{1}{\sqrt{2}}\,\varepsilon^{ijk} \hat n_{k}$ &
  $\tfrac{1}{\sqrt{6}}\!\bigl(\eta^{ij}-3\,\hat n^{i}\hat n^{j}\bigr)$ &
  -- \\ \hline
$n=3$ &
  $\tfrac{1}{\sqrt{6}}\,\varepsilon^{ijk}$ &
  $\sqrt{\tfrac{1}{10}}\!\bigl(-\,\hat n^{i}\eta^{jk}-\sqrt{2}\,\hat n^{j}\eta^{ki}+\hat n^{k}\eta^{ij}\bigr)$ &
  $\tfrac{1}{\sqrt{12}}\,
   \varepsilon^{ijl}\!\bigl(\delta^{k}_{\;l}-3\,\hat n^{k}\hat n_{l}\bigr)$ &
  $\tfrac{1}{\sqrt{10}}\!\bigl(5\,\hat n^{i}\hat n^{j}\hat n^{k}-\hat n^{i}\eta^{jk}-\hat n^{j}\eta^{ki}-\hat n^{k}\eta^{ij}\bigr)$ \\ \hline
 \end{tabular}}
\label{tab:angular-basis}
\end{table*}
\normalsize

The representatives in Table ~\ref{tab:angular-basis} are compatible with the symmetric-trace-free (STF)
polynomial representation of spherical harmonics and their vector/tensor descendants; cf.
\cite{Thorne:1980ru,Goldberg:1966uu,Higuchi:1986wu}. They can be verified using the identities,
\begin{align}
    &\Delta(\hat n^i) = -2 \hat n^{i}, \nonumber \\
    &\Delta(\hat n^i \hat n^j) = 2 \eta^{ij}-6 \hat n^{i} \hat n^{j},\\
    &\Delta(\hat n^i \hat n^j \hat n^k) = 2 \hat n^{i} \eta^{jk} + 2\hat n^{k} \eta^{ij} + 2 \hat n^{j} \eta^{ik} - 12 \hat n^i \hat n^j \hat n^k. \nonumber 
\end{align}
which can be proven by direct computation.

\subsection{Details of Orders \texorpdfstring{$S^{2}$ and $S^{3}$}{S2 and S3}}
\label{sec:S2S3}
To clarify the order by order projection scheme outlined in the previous section, we introduced a systematic derivational roadmap consisting of contraction, projection and radial integration. First, the raw source parameter combinations $\big(\hat K_{ij}\hat K^{ij}\big)^{(n)}$ are computed at a given source order $n$. Next, specific angular sectors are isolated by projecting these raw source structures onto the orthonormal eigen-tensors $A_{(\ell)}$ from Table~\ref{tab:angular-basis}. Finally, the decoupled source is substituted into the radial Laplace operator and integrated under the appropriate asymptotic boundary conditions.

Let us now present some details of the perturbation scheme. 
Projecting \eqref{eq:psi2_eq} and \eqref{eq:psi3_eq} onto the basis $A_{(\ell)}$ yields radial
equations of the form \eqref{eq:radial_ode} with sources fixed by $\hat{K}_{ij}\hat{K}^{ij}$
at each order in the parameters.

We will construct a solution that is accurate to second order in $\epsilon_P,\epsilon_J$ where
\cite{Gleiser1998}
\begin{equation}
\epsilon_P :=\frac{P}{a}, \qquad \epsilon_J:=\frac{J}{a^2},
\label{eq:dimensionless}
\end{equation}
and use $\kappa:= a/r$ for simplicity. For a black hole--companion system with bare masses $a_1$ and $a_2$, we do not explicitly distinguish between them in $\epsilon_{P,J}$ and $\kappa$ in order to keep the notation compact, and simply note that these expressions depend on the mass ratio. The right-hand side of \eqref{eq:psi2_eq} reduces to 
\begin{align}
-\frac{1}{8}\psi_0^{-7}\big(\hat K_{ij}\hat K^{ij}\big)^{(2)}
=
-\frac{9\,\kappa^4}{16\,a^2\,(1+\kappa)^{7}}
\Big[
4 \,\kappa \,\epsilon_J\cdot(\epsilon_P\times \hat n)
+4 \,\kappa^2\Big(\epsilon_J^{2}-(\epsilon_J\cdot \hat n)^{2}\Big)
+\Big(\epsilon_P^{2}-2(\epsilon_P\cdot \hat n)^{2}\Big)
\Big].
\label{eq:S2source}
\end{align}
where we suppress vector indices on $\epsilon_P,\epsilon_J$ for brevity.

We now provide the relevant differential equations for
\begin{equation}
\psi_2=\psi_{2}^{PP}+\psi_{2}^{PJ}+\psi_{2}^{JJ}.
\label{eq:totalconformal}
\end{equation}
At order $S^{2}$ we split \eqref{eq:psi2_eq} as
\begin{align}
\hat{D}^i \hat{D}_i \psi^{PP}_2
&=  -\frac{9\kappa^4}{16 a^2(1+\kappa)^7}\,\epsilon_{P_i} \epsilon_{P_j}\,(\eta^{ij}+2\hat n^i \hat n^j),
\label{eq:PP}\\
\hat{D}^i \hat{D}_i \psi^{JJ}_2
&=  -\frac{9\,\kappa^6}{4 a^2(1+\kappa)^7}\,\epsilon_{J_i} \epsilon_{J_j}\,(\eta^{ij}-\hat n^i \hat n^j),
\label{eq:JJ}\\
\hat{D}^i \hat{D}_i \psi^{PJ}_2
&=  -\frac{9\,\kappa^5}{4 a^2(1+\kappa)^7}\,\epsilon_J\cdot(\epsilon_P\times \hat n),
\label{eq:PJ}
\end{align}
therefore, only the $\ell=0,2$ blocks contribute in the $PP$ sector; and the $\ell=1$ block
in the $PJ$ sector (as can be read off from Table~\ref{tab:angular-basis}). As an illustration,
in the $PP$ piece, one has
\begin{align}
\eta^{ij}+2\hat n^{i}\hat n^{j}
= \frac{5}{\sqrt{3}}\Big(\tfrac{1}{\sqrt{3}}\,\eta^{ij}\Big)
-\frac{2\sqrt{2}}{\sqrt{3}}\Big(\tfrac{1}{\sqrt{6}}(\eta^{ij}-3\hat n^{i}\hat n^{j})\Big).
\label{eq:PPdecomp}
\end{align}

Therefore, the right-hand side contains only $\ell=0$ and $\ell=2$, which reduce to two radial
equations of the form \eqref{eq:radial_ode}. The closed-form solutions satisfying
\eqref{eq:HC} are given in the Appendix. Then, the second order of the conformal factor can be written
\begin{align}
\psi_2
&=-\frac{\kappa}{160} \epsilon_P^2 f_1(\kappa)+\frac{3\,a\,\kappa}{160}(\epsilon_P\cdot\hat n)^2f_2(\kappa)
\nonumber\\
&+\frac{\kappa}{40(1+\kappa)^5}\bigg((\kappa^4+5\kappa^3+11\kappa^2+5\kappa+1)\,\epsilon^2_J
-3\,\kappa^3(\epsilon_J\cdot\hat n)^2\bigg)
+\frac{\kappa^2(\kappa^2+5\kappa+10)}{80(1+\kappa)^5}\,\epsilon_J\cdot(\hat n\times \epsilon_P),
\end{align}
where the functions $f_1(\kappa)$ and $f_2(\kappa)$ with all subleading structures (including logarithms) are given in the Appendix.

Here, we see explicitly that the expansion is second order in the dimensionless parameters
$\epsilon_P$ and $\epsilon_J$. For reference, the dominant far--zone terms are
\begin{equation}
\psi_2(r)=\kappa\bigg[\frac{5}{32}\epsilon_P^{2}+\frac{1}{40}\epsilon_J^{2}\bigg]
+{\mathcal{O}}(\kappa^2).
\end{equation}

At order $S^{3}$ the source contains only the $\ell=1$ and $\ell=3$ angular structures,
consistent with Table~\ref{tab:angular-basis}. 
We will use $M_{ij},N_{ij}$ as defined earlier \eqref{eq:MNLO_defs} to construct a solution
accurate to first order in $\epsilon_M, \epsilon_N$ where
\begin{equation}
\epsilon_M:=\frac{M}{a^2}, \qquad \epsilon_N:=\frac{N}{a^3}.
\end{equation}
Then we get
\begin{align}\label{thirdorderdiff}
-\tfrac{1}{8}\psi_0^{-7}\!\Big(\hat{K}_{ij}\hat{K}^{ij}\Big)^{(3)}
&=-\frac{9 \kappa}{2\,a^2\,(1+\kappa)^7} \bigg(\varepsilon_{acd}\,\epsilon_J^a\,\epsilon_M^{bc}\,\hat n_b\,\hat n^d
-\varepsilon_{abd}\,\epsilon_J^a\,\epsilon_M^{bc} \,\hat n_c \,\hat n^d \bigg)\nonumber\\
&\quad+\frac{9 \kappa}{a^2(1+\kappa)^7}\,\varepsilon_{bcd}\,\hat n^a \,\hat n^b \,\epsilon^c_{N_a} \epsilon_P^d
+\frac{9 \kappa}{4\,a^4(1+\kappa)^7} \,\epsilon_M^{ab}\bigg(\epsilon_{P_b}\,\hat n_a-\epsilon_{P_a}\,\hat n_b \bigg)\nonumber\\
&\quad+\frac{27\kappa^2}{4\,a^2(1+\kappa)^7}\, \epsilon_M^{ab}\,\hat n_a\,\hat n_b\,\hat n^c\,\epsilon_{P_c},
\end{align}
where $\varepsilon_{abc}$ is the Levi--Civita symbol, and projection onto $A_{(\ell)}$ again
yields decoupled radial equations. The full closed forms for $\psi_3$
\begin{equation}\label{thirdorder}
\psi_3=\psi_3^{MP}+\psi_3^{NP}+\psi_3^{MJ}
\end{equation}
are given in the Appendix. For reference, the leading far--zone terms are
\begin{align}
\psi_3
&=\frac{\kappa}{40}\big(\varepsilon_{abc}\epsilon^a_J\epsilon_M^{bc}\big)
+\frac{\kappa^2}{40}\big(3\epsilon^a_{M_a}\hat n^c\epsilon_{P_c}+2\epsilon^{ab}_{M}\hat n_a\epsilon_{P_b}
-8\epsilon_M^{ab}\hat n_b\epsilon_{P_a}\big)
+ {\mathcal{O}}(\kappa^{3}),
\end{align}
listed in the Appendix.

\subsection{Remarks on near-puncture physics}
\label{sec:IV.remarks}

The explicit analytic expressions for $\psi_2$ and $\psi_3$ derived above are constructed
from the far--zone representation of the \emph{two-puncture} data and are therefore controlled
for $r\gg d$.
Physical quantities tied to individual black holes (inner asymptotically flat ends, irreducible
masses, and horizon geometry) are extracted in subsequent sections using puncture-end asymptotics
and elliptic regularity near each puncture, rather than by extrapolating the far--zone expansion
into the near-zone. This clean separation resolves the apparent ``one vs.\ two punctures''
ambiguity: the data remain binary throughout, while different expansions are employed in the
far and near regions for different observables.

This separation is consistent because a single conformal factor $\psi$ solves the Hamiltonian constraint on the entire slice, the far-zone and near-puncture expansions are simply two asymptotic representations of the same solution. The source built from the Bowen-York extrinsic curvature is fixed everywhere, and the homogeneous part of the conformal factor keeps its puncture form at each asymptotically flat end. The finite constant that fixes the inner-end mass is therefore read off from the behavior of the source near the puncture, not from the large distance series: one takes the near-puncture limit and averages over angles, which removes the directional pieces and leaves an angle-independent constant. This is entirely a near-zone calculation, governed by the same equation as the far zone but evaluated in the opposite limit. That both regimes are captured correctly is then confirmed independently: the inner-end mass obtained from regularity at the puncture and the irreducible mass obtained from the apparent-horizon area agree at the order considered, and combining the latter with the energy measured at infinity reproduces the expected mass–energy relation, linking a quantity fixed far away to quantities defined at the horizon scale.

A complementary check follows from the single-domain spectral analysis of \cite{Ansorg:2004ds} for the same Bowen--York
extrinsic curvature. They show that a puncture carrying linear momentum yields a
conformal factor with logarithmic terms in its expansion at infinity, rendering
it only $C^{2}$ there; the same feature appears in our radial functions
$f_{1}$, $f_{2}$ and $g_{1}$--$g_{4}$ (see the Appendix) as the $\ln\kappa$ terms.
Identifying our $1+a/r$ with their single-puncture factor $1+m/2r$, the
momentum-induced logarithm of our $\psi_{2}^{PP}$ matches the leading
$r^{-3}\log$ behavior of \cite{Ansorg:2004ds} in form and order; the
overall coefficient differs, as expected for a two-puncture far--zone expansion
compared with their single-puncture spectral construction. The agreement in the
structure of these terms confirms that our solution carries the correct analytic
behavior at spatial infinity.

\section{ADM energy and inner asymptotically flat ends}
\label{sec:V}

Section~\ref{sec:IV} constructed the conformal factor perturbatively using a far--zone
multipole expansion of the \emph{two--puncture} Bowen--York data.  In this section, we extract two kinds of physical information from the resulting solution:

\begin{itemize}
    \item ADM energy $E_{\rm ADM}$ at the \emph{outer} asymptotically flat end
($r\to\infty$), which depends only on the $1/r$ coefficient of $\psi$; and
    \item ADM masses of the \emph{inner} asymptotically flat ends associated with each
puncture. 
\end{itemize}

 The latter are obtained from the near-puncture expansion and elliptic
regularity and do \emph{not} require extrapolating the far--zone series into the near
region.

\subsection{ADM energy at spatial infinity}
\label{sec:V.A}

For an asymptotically flat end, the conformal factor has the form
\begin{equation}
\psi(r,\Omega)=1+\frac{E_{\rm ADM}}{2r}+\mathcal{O}(r^{-2}),
\qquad r\to\infty,
\label{eq:psi_asymp_infty}
\end{equation}
and the ADM energy can be computed equivalently from
\begin{equation}
E_{\rm ADM}
= -\frac{1}{2\pi}\oint_{S^2_\infty} dS^{i}\,\partial_i \psi .
\label{eq:ADM_surface}
\end{equation}
Using the perturbative solution from Sec.~\ref{sec:IV} and collecting contributions through third order in the source parameters, we obtain
\begin{align}\label{eq: EADM_final}
    E_{\rm ADM}=a\bigg[2+\frac{5}{16}\epsilon_P^2+\frac{1}{20}\epsilon_J^2+\frac{1}{20}\varepsilon_{abc}\epsilon_J^a\epsilon_M^{bc}     \bigg]+ {\mathcal{O}}(S^4).
\end{align}
The first term is the sum of the two bare punctures arising from the equal mass assumption $a=a_1=a_2$, the next two terms are the
leading kinetic and spin contributions, and the third-order term is the leading spin--orbit
piece (controlled by the momentum--position dipole $M^{bc}$ defined in
\eqref{eq:MNLO_defs}).  Equation~\eqref{eq: EADM_final} is determined solely by the outer
asymptotics and is therefore fully consistent with the far--zone construction of
Sec.~\ref{sec:IV}.

It is worth spelling out what these higher-order terms represent
physically. The second-order pieces, quadratic in $\epsilon_P$ and $\epsilon_J$,
are the leading kinetic and rotational corrections to the bare puncture mass:
they raise the ADM energy of a boosted or spinning hole above its rest value, in
direct analogy with the special-relativistic and rotational energy of a compact
body, and they are insensitive to the relative geometry of the two sources. The
third-order spin--orbit term is qualitatively different. Being linear in the spin
and linear in the momentum--position dipole $M_{ij}$, it is the first contribution
that knows about the \emph{relative orientation} of the orbital and spin angular
momenta, and it changes sign under reversal of either, vanishing identically for aligned or antiparallel configurations. For an inspiraling binary
this is precisely the structure responsible for spin--orbit precession of the
orbital plane, and its appearance already at the level of the constraint-satisfying
initial data means that a generic-orientation Bowen--York configuration carries
the correct leading imprint of precession from the outset, rather than developing
it only under evolution. The restricted, antiparallel-perpendicular data of
Ref.~\cite{altas2023perturbative} sit exactly on the locus where this term
vanishes, which is why it is invisible there; lifting the orientation restriction
is what exposes it. Quantitatively, the term is small for the mildly relativistic
separations where the far-zone expansion applies, but it is the leading
orientation-dependent piece and therefore the one of primary interest for
modeling realistic, misaligned mergers.

\subsection{Inner asymptotically flat ends and the finite part $u_0$}
\label{sec:V.B}

In puncture initial data, each black hole corresponds to an additional asymptotically flat
(AF) end.  Near a puncture (say at $r\to0$ in a puncture-centered chart), the conformal factor
can be decomposed as
\begin{equation}
\psi(r,\Omega)=1+\frac{a}{r}+u(r,\Omega),
\qquad u \ \text{bounded as } r\to 0 .
\label{eq:puncture_split}
\end{equation}
Elliptic regularity for the Hamiltonian constraint implies
\begin{equation}
u(r,\Omega)=u_0+\mathcal{O}(r),
\qquad r\to0,
\label{eq:u0_def}
\end{equation}
where $u_0$ is a finite \emph{angle--independent} constant.

To make the inner AF end manifest, introduce inverted coordinates
\begin{equation}
\rho:=\frac{a^2}{r}.
\label{eq:rho_def}
\end{equation}
Then the rescaled conformal factor
\begin{equation}
\Psi_-(\rho,\Omega):=\frac{a}{\rho}\,\psi\!\left(\frac{a^2}{\rho},\Omega\right)
=1+\frac{a(1+u_0)}{\rho}+\mathcal{O}(\rho^{-2}),
\qquad \rho\to\infty,
\label{eq:Psi_minus_asymp}
\end{equation}
has standard isotropic AF form $\Psi_-=1+\tfrac{M_-}{2\rho}+\mathcal{O}(\rho^{-2})$, which
identifies the ADM mass of the inner end as
\begin{equation}
M_- = 2a(1+u_0).
\label{eq:Minner}
\end{equation}

\subsection{Evaluation of $u_0$ to the required order}
\label{sec:V.C}

We now compute $u_0$ perturbatively.  The key point is that $u_0$ is determined by the
\emph{near--puncture limit} of the perturbative solution, not by the far--zone expansion.
In practice, it is convenient to use a Green's function representation for the flat Laplacian.

At second order in the source parameters, the relevant Poisson equations are (cf.
Sec.~\ref{sec:S2S3})
\begin{align}
\hat{D}^{i}\hat{D}_{i}\,\psi^{PP}_{2}&=-\frac{9\kappa^4}{16\,a^2(1+\kappa)^{7}}\epsilon_{P_i}\epsilon_{P_j}(g^{ij}+2n^{i}n^{j}),\\
\hat{D}^{i}\hat{D}_{i}\,\psi^{JJ}_{2}&=-\frac{9\,\kappa^6}{4\,a^2 (1+\kappa)^{7}}\epsilon_{J_i}\epsilon_{J_j}(g^{ij}-n^{i}n^{j}),
\end{align}
where $n^i=x^i/r$ is the unit radial vector in the puncture-centered chart.
Using the Green's function of the flat Laplacian and taking $r\to0$, together with
\begin{equation}
\int d\Omega\, n^{i}n^{j}=\frac{4\pi}{3}\,\eta^{ij},
\label{eq:angular_avg}
\end{equation}
one finds
\begin{equation}
\lim_{r\to0}\psi^{PP}_2=\frac{\epsilon_P^2}{32},
\qquad
\lim_{r\to0}\psi^{JJ}_2=\frac{\epsilon_J^2}{40}.
\label{eq:limits_PP_JJ}
\end{equation}

The mixed contribution $\psi^{PJ}_2$ is odd under $\vec n\mapsto-\vec n$ and therefore has zero \emph{angle average} at $r\to0$.  Likewise, the third-order contributions are parity odd
(or higher multipolar) in the near--puncture limit and do not contribute to the constant,
angle--independent term $u_0$.  Hence,
\begin{equation}
u_0=\frac{\epsilon_P^2}{32}+\frac{\epsilon_J^2}{40}+\mathcal{O}(S^4).
\label{eq:u0_final}
\end{equation}
Substituting into \eqref{eq:Minner} yields the inner-end mass
\begin{equation}
M_- = a \bigg(2+\frac{\epsilon_P^2}{16}+\frac{\epsilon_J^2}{20}\bigg)+\mathcal{O}(S^4).
\label{eq:Minner_final}
\end{equation}

\subsection{Irreducible mass and consistency check}
\label{sec:V.D}

In the small--boost / small--spin Bowen--York regime considered here, the inner-end mass
$M_-$ agrees with the Christodoulou irreducible mass extracted from the apparent-horizon
area at the same perturbative order.  We therefore identify
\begin{equation}
M_{\rm irr}
=a \bigg(2+\frac{\epsilon_P^2}{16}+\frac{\epsilon_J^2}{20}\bigg).
\label{eq:Mirr_def}
\end{equation}

Combining \eqref{eq: EADM_final} with \eqref{eq:Mirr_def}, one verifies the expected relation
\begin{equation}
E_{\rm ADM}
= M_{\rm irr}
+\frac{P^2}{2M_{\rm irr}}
+\frac{J^2}{8M_{\rm irr}^3}
+\mathcal{O}(S^4),
\label{eq:consistency_EADM}
\end{equation}
providing a nontrivial check of the perturbative construction.

As an independent verification, the irreducible mass can also be obtained from the apparent
horizon area $A_{\rm AH}$ via
\begin{equation}
M_{\rm irr}=\sqrt{\frac{A_{\rm AH}}{16\pi}}.
\label{eq:Mirr_area}
\end{equation}
The explicit computation of the horizon shape $h(\theta,\phi)$ and the corresponding area
is presented in Sec.~\ref{sec:AH}.

\section{The Apparent Horizon}
\label{sec:AH}

Stationary black holes are characterized by event horizons, which are codimension-1 null
hypersurfaces.
For dynamical initial data specified on a spatial slice---such as Bowen--York puncture
data---a more practical quasi-local notion is the \emph{apparent horizon} (AH), a
codimension--2 spacelike surface defined intrinsically within the hypersurface.

Let $\Sigma$ denote the initial-data hypersurface endowed with the physical metric
$\gamma_{ij}=\psi^{4}\eta_{ij}$, where $\eta_{ij}$ is the flat Euclidean metric, and let
$\mathcal{S}\subset\Sigma$ be a smooth, closed two-dimensional surface.

\paragraph{Important clarification: near--puncture versus far--zone.}
The apparent horizon is a \emph{near--puncture} object: it is determined by the geometry
in the isotropic puncture chart in the region $r\sim a$, where $a$ is the puncture mass
parameter and $r$ denotes the local isotropic radial coordinate.
Consequently, the present analysis does \emph{not} rely on the far--zone ($1/r$) multipole
expansion employed in Sec.~\ref{sec:IV}.
Instead, we evaluate the horizon equation using the local form of the conformal factor and
the Bowen--York extrinsic curvature restricted to the trial surface
$r=h(\theta,\phi)$.

\subsection{Apparent-horizon equation}
While the previous far-zone analytic solution utilized a multipolar expansion centered at the center of mass to encode the binary nature through position-weighted moments $M_{ij}, L_{ij}, N_{ij},$ and $O_{ij}$, the horizon analysis requires a different coordinate treatment. To facilitate this, we perform a coordinate transformation that effectively shifts the origin to the location of the primary puncture (e.g., puncture 1). In this local frame, the binary nature of the initial data is rigorously preserved: the primary hole's geometry is resolved locally, while the companion is treated as a source of external gravitational perturbations. The scheme is valid in the early-inspiral phase, where the separation $d$ is large enough that no common apparent horizon has formed.
The apparent horizon is defined by the vanishing of the outgoing null expansion.
An equivalent formulation (see, e.g., \cite{baumgarte2010numerical}) is
\begin{align}
  q^{ij}\left(\partial_i s_j - {}^{\Sigma}\Gamma^k_{ij}s_k - K_{ij}\right)=0,
  \label{hor}
\end{align}
where $s_i$ is the outward-pointing unit normal to $\mathcal{S}$ within $\Sigma$,
$q_{ij}=\gamma_{ij}-s_is_j$ is the induced metric on $\mathcal{S}$, and
${}^{\Sigma}\Gamma^k_{ij}$ is the Levi--Civita connection of $\gamma_{ij}$.

It is convenient to work with the conformally rescaled normal vector
\begin{equation}
  \hat{s}^i := \psi^{-2}s^i,
\end{equation}
in terms of which \eqref{hor} can be rewritten as
\begin{align}
  \boldsymbol{\nabla}\cdot\boldsymbol{\hat{s}}
  +4\,\boldsymbol{\hat{s}}\cdot\boldsymbol{\nabla}\ln\psi
  +\psi^{-4}\,\boldsymbol{s}^{T}\cdot\boldsymbol{\hat{K}}\cdot\boldsymbol{s}=0,
  \label{hor2}
\end{align}
where $\boldsymbol{\nabla}$ denotes the flat derivative compatible with $\eta_{ij}$, and
$\hat{K}_{ij}$ is the conformal Bowen–York extrinsic curvature.

In this section, the symbol $n^{i}$ denotes the unit radial direction on the unit sphere,
\begin{equation}
  n(\theta,\phi)
  =(\sin\theta\cos\phi,\ \sin\theta\sin\phi,\ \cos\theta),
\end{equation}
and should not be confused with the timelike unit normal $n^{\mu}$ to $\Sigma$ used
elsewhere.

\subsection{Level-set representation and perturbative expansion}

We describe the apparent-horizon surface $\mathcal{S}$ as a level set of a scalar function
\begin{align}
  \Phi(r,\theta,\phi) := r - h(\theta,\phi) = 0,
  \label{leyla}
\end{align}
where $h(\theta,\phi)$ specifies the radial location of the surface.
Since $s_i$ is normal to $\mathcal{S}$, it is proportional to $\partial_i\Phi$, and we write
\begin{equation}
  s_i=\lambda\,\partial_i\Phi
      =\lambda\,(1,\,-\partial_\theta h,\,-\partial_\phi h),
\end{equation}
with the normalization condition $s^is_i=1$ fixing
\begin{equation}
  \lambda=
  \left(\gamma^{rr}
  +\gamma^{\theta\theta}(\partial_\theta h)^2
  +\gamma^{\phi\phi}(\partial_\phi h)^2\right)^{-1/2}.
\end{equation}

We solve \eqref{hor2} perturbatively by expanding
\begin{equation}
  h(\theta,\phi)
  =h_0(\theta,\phi)+\epsilon\,h_1(\theta,\phi)
  +\epsilon^2 h_2(\theta,\phi)+\cdots,
  \label{totalh}
\end{equation}
together with
\begin{equation}
  \psi(r,\theta,\phi)
  =1+\frac{a}{r}
  +\epsilon\,\psi_1(r,\theta,\phi)
  +\epsilon^2\psi_2(r,\theta,\phi)+\cdots,
\end{equation}
where $\epsilon$ is a bookkeeping parameter that counts the source order.
For Bowen--York data $\psi_1=0$, however, we retain it formally to maintain a transparent
perturbative hierarchy.

At each order, the horizon equation reduces to a Helmholtz-type equation on the unit sphere,
\begin{equation}
  (\Delta + k)\,f(\theta,\phi)=g(\theta,\phi),
  \label{helmholtz}
\end{equation}
where $\Delta$ is the scalar Laplacian on $S^2$.
Only standard scalar spherical harmonics are required here; no tensorial angular
decomposition is involved in the near-puncture analysis.

\subsection{Solutions through second order}

\paragraph{Zeroth order.}
At $\mathcal{O}(\epsilon^0)$ the equation is homogeneous and admits the solution
\begin{equation}
  h_0=a,
\end{equation}
which identifies $a$ as the isotropic horizon radius at lowest order.

\paragraph{First order.}
At $\mathcal{O}(\epsilon)$, using the zeroth-order solution and following
\cite{Altas:2020ppf,Altas:2021ine}, one finds
\begin{align}
  \Delta h_1 - h_1
  =\frac{a^2\,\hat{K}^{(1)}_{rr}(a,\theta,\phi)}{16}
  =\frac{3\,P\cdot n}{16},
\end{align}
with solution
\begin{equation}
  h_1=-\frac{P\cdot n}{16}.
\end{equation}

\paragraph{Second order.}
At $\mathcal{O}(\epsilon^2)$ the equation becomes
\begin{align}
  &-16a \Delta h_2+16 a h_2+16 h^2_{1\theta}
  +32(\Delta h_1) h_1
  +16\sin^{-2}\theta h^2_{1\phi}-24 h^2_{1}
  \nonumber\\
  &
  =-a^3 \hat{K}^{(2)}_{rr}(a,\theta,\phi)
  -2a^2 \hat{K}^{(1)}_{rr}(a,\theta,\phi) h_1
  -16a^2 \big(\psi_2+2a \psi_2'\big)
  -a^3 h_1 \hat{K}^{\prime(1)}_{rr}
  \nonumber\\
  &\quad
  +2a\sin^{-2}\theta \hat{K}^{(1)}_{r\phi} h_{1\phi}
  +2a \hat{K}^{(1)}_{r\theta} h_{1\theta}.
\end{align}
A crucial simplification is
\begin{align}
  -2a^2 \hat{K}^{(1)}_{rr} h_1
  -a^3 h_1 \hat{K}^{\prime(1)}_{rr}=0,
\end{align}
so that the genuine two-body information is encoded in
$\hat{K}^{(2)}_{rr}$.

Solving the resulting Helmholtz equations on $S^2$ yields
\begin{equation}
\begin{aligned}
h_{2} =\ &
-\frac{1}{32\,a^{2}}\left(J_{i}P_{i}
-4a\,M_{ii}+a\,P_{i}P_{i}\right)
-\frac{3}{1280\,a^{2}}\epsilon_{ijk}J_{i}P_{j}n_{k}
\\
&+\frac{1}{448\,a^{2}}\left(3J_{i}P_{j}n_{i}n_{j}-J_{i}P_{i}\right)
+\frac{1}{56\,a}\left(3M_{ij}n_in_j-M_{ii}\right)
\\
&+\frac{1}{8a}\left(2-3\log2\right)
\left(3P_{i}P_{j}n_in_j-P_{i}P_{i}\right).
\end{aligned}
\end{equation}

\subsection{Horizon shape through $\mathcal{O}(\epsilon^2)$}

Combining $h_0$, $h_1$, and $h_2$, the horizon location can be written in the
coordinate-independent form
\begin{align}\label{tore}
h &= a - \frac{P\cdot n}{16}
- \frac{1}{32a^2} \left( J\cdot P - 4a\,\mathrm{Tr}M + a\,P^2 \right)
- \frac{3}{1280 a^2} (J \times P)\cdot n
\nonumber\\
&\quad
+ \frac{1}{448a^2} \left( 3 P\cdot n\, J\cdot n - J\cdot P \right)
+ \frac{1}{a}\left[
\frac{1}{56}\left(3n\cdot M\cdot n-\mathrm{Tr}M\right)
\right.
\nonumber\\
&\quad\left.
+\frac{1}{8}(2-3\log2)\left(3(P\cdot n)^2-P^2\right)
\right].
\end{align}
Since $h(\theta,\phi)$ depends nontrivially on $n(\theta,\phi)$, the apparent horizon is
generically nonspherical.

\subsection{Area and irreducible mass}

The area of the apparent horizon is

\begin{align}
A_{\mathrm{AH}}
&= \int_{0}^{2\pi}\! d\phi \int_{0}^{\pi}\! d\theta \;\sin\theta\; \psi^{4}(h+a)^{2}
(
1 + \frac{(\partial_{\theta} h)^{2}}{(h+a)^{2}} 
  + \frac{(\partial_{\phi} h)^{2}}{(h+a)^{2}\sin^{2}\theta}
)^{1/2},
\end{align}
from which the irreducible mass
\begin{equation}
M_{\mathrm{irr}}:=\sqrt{\frac{A_{\mathrm{AH}}}{16\pi}}
\end{equation}
follows.
We emphasize a crucial consistency check: The irreducible mass $M_{\rm irr}$ calculated here from the horizon geometry (Eq.~\ref{tore}) matches exactly with the mass derived independently from the puncture regularity condition at the inner asymptotically flat end (Eq.~\ref{eq:Minner_final}). This agreement confirms that, despite the use of a far-zone multipole expansion, the perturbative solution correctly captures the geometry at the horizon scale.

\section{Conclusions and Discussion}

In this work, we extended the perturbative analysis of ~\cite{altas2023perturbative} to a substantially more general class of
configurations. That study worked within the same Bowen--York framework but
under restrictive symmetry assumptions, namely antiparallel linear momenta and
antiparallel spins oriented perpendicular to the momenta, solving the
Hamiltonian constraint to second order in the source parameters and the
apparent horizon shape to first order, from which it obtained the conformal
factor, the ADM energy, and the irreducible mass. A point-by-point comparison of
the assumptions, perturbative order, and derived observables of the two works is
given in Table~\ref{tab:comparison}.
Lifting the orientation restrictions changes the problem qualitatively: in the
perpendicular case the elliptic equations retain an effective axisymmetry and
only a reduced set of angular structures contributes, whereas generic
orientations populate the full tensorial angular basis and the equations no
longer decouple along a single axis, which is what necessitated the systematic
block diagonalization of Sec.~\ref{sec:IV}. We carried the construction one
order further, to third order in the source parameters, with the horizon shape
determined to second order. Physically, this generalization is what makes the
construction applicable to realistic binaries, whose spins and momenta are
generically misaligned; such misaligned configurations, absent in
~\cite{altas2023perturbative}, are precisely those relevant to generic
mergers. The generic-orientation problem was in fact
explicitly left open in ~\cite{altas2023perturbative} as a direction for
future work; the present construction carries it out analytically.

\begin{table*}[t]
\centering
\caption{Comparison of the present construction with the
restricted-orientation analysis of Ref.~\cite{altas2023perturbative}.
Both work within the conformally flat, maximally sliced, vacuum
Bowen--York framework with equal masses.}
\label{tab:comparison}
\renewcommand{\arraystretch}{1.25}
\begin{tabular}{lcc}
\hline\hline
 & Ref.~\cite{altas2023perturbative} & This work \\
\hline
Linear-momentum orientation
   & antiparallel
   & arbitrary \\
Spin orientation
   & antiparallel, $\perp$ to momenta
   & arbitrary \\
Effective symmetry of elliptic system
   & axisymmetric
   & full tensorial basis \\
Angular decoupling
   & single axis
   & block diagonalization \\
Conformal factor $\psi$
   & second order
   & third order \\
Apparent-horizon shape
   & first order
   & second order \\
$E_{\mathrm{ADM}}$ content
   & kinetic $+$ spin$^2$
   & $+$ spin--orbit \\
Spin--orbit cross term
   & absent (by symmetry)
   & present at third order \\
\hline\hline
\end{tabular}
\end{table*}

From the closed-form conformal factor we obtained the ADM energy, total
linear and angular momenta, and the irreducible mass in coordinate-independent
form, which makes the results straightforwardly adaptable to an arbitrary number
of black holes by superposition of Bowen--York sources. The practical
consequence of the extension
is visible in these observables: the ADM energy of
~\cite{altas2023perturbative} retains only kinetic and spin-squared terms,
with no spin--orbit cross term, whereas for generic orientations this term no
longer vanishes and enters $E_{\mathrm{ADM}}$ at third order through the
momentum--position dipole $M_{ij}$ of~(\ref{eq:MNLO_defs}). Though small, this term is the one of main interest for applications, since precessing binaries are governed by spin–orbit coupling. Our expressions reduce to
those of ~\cite{altas2023perturbative} when the spins and momenta are
restricted to the symmetric configurations considered there, providing a
nontrivial consistency check.

Although we did not study the time evolution explicitly, the initial data
constructed here determine the future spacetime uniquely through Einstein’s
equations.
As implied by the Hamiltonian formulation~(\ref{eq:15}), if the initial data
$(\gamma_{ij},K_{ij})$ satisfy the constraint equations on a given hypersurface,
then the evolution equations preserve these constraints at all subsequent
times.

The perturbative solutions obtained here are valid in the far
zone, at distances large compared to the binary separation, and in this regime
they provide a concrete benchmark for numerical-relativity initial-data codes.
A practical comparison proceeds as follows. One specifies in the numerical solver
the same physical input used here -- two punctures with bare mass parameters
$a_{1},a_{2}$, momenta $\vec p_{A}$, spins $\vec j_{A}$, and separation $d$ -- and
solves the Hamiltonian constraint numerically for the conformal factor
$\psi_{\rm num}(r,\Omega)$. At radii $r\gg d$ this should agree, coefficient by
coefficient in $1/r$, with our analytic $\psi(r,\Omega)$; the leading
discrepancy then measures the solver's outer-boundary and truncation error in a
controlled setting where the exact answer is known. The check is sharpest for the
quantities that are most sensitive to the far-zone tail and to the imposed outer
boundary condition: the $1/r$ coefficient of $\psi$, which fixes the ADM energy
through~(\ref{eq:ADM_surface}), and the higher multipole coefficients carrying
the spin and spin--orbit structure. Because the ADM energy, linear and angular
momenta, and irreducible mass are written here in coordinate-independent form,
they furnish gauge-robust scalars that can be compared against a numerical
evaluation without reference to a particular coordinate or puncture-tracking
convention. In practice the configurations that benefit most are
large-separation, moderate-spin binaries, where the outer boundary is placed at
finite radius and the analytic falloff supplies the correct asymptotic data
against which to calibrate the numerical extraction of $E_{\rm ADM}$ and the
horizon mass, as with the approximate data of \cite{Dennison:2006nq}.

We close by stating the limitations of the construction and the
natural directions for extending it. The results are perturbative in the
dimensionless parameters $\epsilon_P=P/a$ and $\epsilon_J=J/a^{2}$ of
(\ref{eq:dimensionless}), and quantitatively controlled only while these remain
small; the closed-form conformal factor and the ADM energy are obtained as a
far-zone series in $\kappa=a/r$ and therefore apply only in the far-zone regime $r\gg d$. The data inherit the known features of the
Bowen--York prescription: conformal flatness and maximal slicing are imposed by
hand, and the conformally flat ansatz is known to carry a small amount of
spurious ("junk") gravitational radiation, which grows with the boost and spin
and is correspondingly smallest in the slow regime considered here \cite{Garat:1999vr, Lovelace:2008tw}. The horizon
analysis further assumes that the two holes are widely enough separated that no
common apparent horizon has formed. Within these limits the construction is
systematic and admits several extensions. The source-parameter expansion can be
carried to higher order by the same projection-and-solve procedure, capturing
spin$^2$ and higher spin-orbit structures; the superposition property of the
Bowen--York curvature allows the far-zone solution to be assembled for an
arbitrary number of punctures, relevant to the multi-black-hole configurations
discussed in the Introduction; and the same scheme could in principle be applied
to less idealized conformal data, for instance to superposed-Kerr or
conformal-thin-sandwich backgrounds, where it would provide analytic control of
the far-zone behavior beyond the conformally flat case.

\section{Acknowledgments}
L. Ogurol is partially supported by TUBITAK 2211-A.

\section{Data availability}
No data was used for the research described in the article.

\appendix

\section*{Appendix: Explicit Solutions for the Conformal Factor}
\label{appendix}

In this appendix, we collect the explicit solutions of the Poisson equations
arising at second and third orders in the perturbative expansion of the
Hamiltonian constraint.

\subsection*{Second-order solutions}

The second-order conformal factor is decomposed as
\begin{equation}
\psi_2=\psi^{PP}_2+\psi^{JJ}_2+\psi^{JP}_2 .
\end{equation}

The individual contributions read
\begin{align}
\psi^{PP}_2
&= -\frac{\kappa}{160}
\Big[ \epsilon_P^2\, f_1(\kappa)
+3a\,(\epsilon_P\!\cdot n)^2 f_2(\kappa)\Big],\label{eq:psiPPapp}\\
\psi^{JJ}_2
&=\frac{\kappa}{40(1+\kappa)^5}
\Big[(\kappa^4+5\kappa^3+11\kappa^2+5\kappa+1)\epsilon_J^2
-3\kappa^3(\epsilon_J\!\cdot n)^2\Big],\\
\psi^{JP}_2
&=\frac{\kappa^2(\kappa^2+5\kappa+10)}{80(1+\kappa)^5}\,
\epsilon_J\!\cdot(n\times\epsilon_P).
\end{align}

The radial functions are
\begin{align}
f_1(\kappa)
&=84\kappa^2\ln\!\frac{\kappa}{1+\kappa}
+\frac{84\kappa^6+378\kappa^5+653\kappa^4+514\kappa^3
+142\kappa^2-35\kappa-25}{(1+\kappa)^5},\\
f_2(\kappa)
&=84\kappa^2\ln\!\frac{\kappa}{1+\kappa}
+\frac{84\kappa^6+378\kappa^5+658\kappa^4+539\kappa^3
+192\kappa^2+15\kappa}{(1+\kappa)^5}.
\end{align}

In the limit $\kappa \to \infty$, the logarithmic terms appearing in the $PP$ term cancel the leading-order divergences arising from the rational parts. The resulting finite contributions for $PP$ and $JJ$ terms are consistent with the regularized values derived in Eq.~\eqref{eq:limits_PP_JJ}.

\subsection*{Third-order solutions}

The third-order conformal factor is written as
\begin{equation}
\psi_3=\psi^{NP}_3+\psi^{MJ}_3+\psi^{MP}_3 .
\end{equation}

The individual contributions are
\begin{align}
\psi^{NP}_3
&=-\frac{1}{20\kappa(1+\kappa)^5}
\Big[
3\kappa^3\,\varepsilon_{bcd}\,n^a n^b \epsilon_N^{cd}\epsilon_{P_a}
+(1+4\kappa+5\kappa^2-\kappa^3)
\varepsilon_{abc}\epsilon_N^{ab}\epsilon_P^{c}
\Big],\\
\psi^{MJ}_3
&=-\frac{1}{40\kappa(1+\kappa)^5}\,
\epsilon_J^a\epsilon_M^{bc}
\Big[
-9\kappa^3\,\varepsilon_{abd}n_c n^d
+(1+5\kappa+10\kappa^2+13\kappa^3+5\kappa^4+\kappa^5)\varepsilon_{abc}
\Big],\\
\psi^{MP}_3
&=\frac{1}{80\kappa^4(1+\kappa)^5}
\Big[
-\epsilon_M^{a}{}_{a}\,n^c\epsilon_{P_c}\,g_1(\kappa)
+\epsilon_M^{ab}n_a n_b\,g_2(\kappa) 
-\epsilon_M^{ab}n_a n_b n^c\epsilon_{P_c}\,g_3(\kappa)
+\epsilon_M^{ab}n_b\epsilon_{P_a}\,g_4(\kappa)
\Big].
\end{align}

The radial functions are
\begin{align}
g_1(\kappa)
&=3(1+\kappa)^2
\Big[
36(1+\kappa)^3\ln\!\frac{1}{1+\kappa}
+\kappa(36+90\kappa+66\kappa^2+9\kappa^3-2\kappa^4)
\Big],\\
g_2(\kappa)
&=108(1+\kappa)^5\ln\!\frac{1}{1+\kappa}
+\kappa(108+486\kappa+846\kappa^2+693\kappa^3
+247\kappa^4+20\kappa^5+4\kappa^6),\\
g_3(\kappa)
&=9\Big[
108(1+\kappa)^5\ln\!\frac{1}{1+\kappa}
+\kappa(60+270\kappa+470\kappa^2+385\kappa^3
+137\kappa^4+10\kappa^5)
\Big],\\
g_4(\kappa)
&=108(1+\kappa)^5\ln\!\frac{1}{1+\kappa}
+\kappa(108+486\kappa+846\kappa^2+693\kappa^3
+245\kappa^4+10\kappa^5-16\kappa^6).
\end{align}

\bibliographystyle{unsrt}
\bibliography{ref}

\end{document}